\documentclass[prb,aps,twocolumn,showpacs,groupedaddress]{revtex4}
\usepackage[dvipdfm]{graphicx}

\begin{document}

\title{Evidences of a consolute critical point in the Phase Separation regime of
La$_{5/8-y}$Pr$_{y}$Ca$_{3/8}$MnO$_{3}$ (y $\sim$0.4) single
crystals}

\author{G. Garbarino}
\author{C. Acha}
\thanks {Also fellow of CONICET of Argentina. Fax: 0054-11-45763357}
\email{acha@df.uba.ar} \affiliation{Laboratorio de Bajas
Temperaturas, Departamento de F\'{\i}sica, FCEyN, Universidad de
Buenos Aires, Ciudad Universitaria, (C1428EHA) Buenos Aires,
Argentina}
\author{P. Levy}
\thanks {Also fellow of CONICET of Argentina.}
\affiliation{Departamento de F\'{\i}sica, CAC, Comisi\'on Nacional
de Energ\'{\i}a At\'omica, Gral Paz 1499, 1650 San Mart\'{\i}n,
Buenos Aires, Argentina}
\author{T. Y. Koo}
\affiliation{Pohang Accelerator Laboratory, Pohang University of
Science and Technology, Pohang, 790-784, S. Korea.}

\author{S-W.Cheong}
\affiliation{Rutgers Center for Emergent Materials and Department of
Physics and Astronomy, Rutgers University, New Jersey, USA}

\date{\today}

\draft

\begin{abstract}

We report on DC and pulsed electric field sensitivity of the
resistance of mixed valent Mn oxide based
La$_{5/8-y}$Pr$_{y}$Ca$_{0.375}$MnO$_{3}$ (y$\sim$0.4) single
crystals as a function of temperature. The low temperature regime of
the resistivity is highly current and voltage dependent. An
irreversible transition from high (HR) to a low resistivity (LR) is
obtained upon the increase of the electric field up to a temperature
dependent critical value ($V_c$). The current-voltage
characteristics in the LR regime as well as the lack of a variation
in the magnetization response when $V_c$ is reached indicate the
formation of a non-single connected filamentary conducting path. The
temperature dependence of $V_c$ indicates the existence of a
consolute point where the conducting and insulating phases produce a
critical behavior as a consequence of their separation.

\end{abstract}

 \maketitle

\section{INTRODUCTION}

A controlled mixture of charge-delocalized ferromagnetic (CD-F) and
charge-ordered, antiferromagnetic (CO-AF) phases can be easily
achieved in manganites that are in the phase separation (PS) regime.
This was indeed observed in previous experiments in hole-doped
manganites where the proportion of the CD-F phase was varied by
changing temperature, electric or magnetic field, grain size or
solely by waiting the evolution of the system to a different phase
distribution~\cite{Asamitsu97,Levy00,Hardy01,Babushkina00,Pandey03}.
Changes of several orders of magnitude were observed, for example,
in the resistivity of the (La,Pr,Ca)MnO$_3$ compound, that were
associated with the variation of the CD-F to CO-AF phase proportion,
which determines the percolation scenario that governs the physics
of this system~\cite{Uehara99}. The study of the properties of the
electric-field-dependent phase distribution may reveal the nature of
some intrinsic attributes of these PS materials, like their
topological phase distribution~\cite{Stank00}, their electrical
transport mechanism~\cite{Markovich05} or their dynamical magnetic
properties~\cite{Ghivelder05}.

In this paper we present the electric field sensitivity of
resistivity and magnetization as a function of temperature in
La$_{5/8-y}$Pr$_{y}$Ca$_{3/8}$MnO$_{3}$ (y$\sim$0.4) single crystals
(LPCMO). We show that the two phase percolation model, where the
CD-F to CO-AF phase proportion increases as the applied electric
field is increased, can only partially explain our results. A
filamentary dielectric breakdown scenario gives instead a better
understanding of the measured properties, where the observed
critical behavior of the breakdown field points out to the existence
of a consolute point related to the separation of the CD-F and the
CO-AF phases.

\section{EXPERIMENTAL}

We have performed resistivity measurements as a function of
temperature and electric field, $\rho(T,V)$, for temperatures in the
4 K$\leq$ T $\leq$ 300 K range,  and electric fields up to 200 V
cm$^{-1}$, in La$_{5/8-y}$Pr$_{y}$Ca$_{3/8}$MnO$_{3}$ (LPCMO,
y$\sim$0.4) single crystals. Details of their synthesis and
characterization can be found elsewhere\cite{Lee02}. Different
contact configurations were used: a four terminal (4W) standard
configuration for constant current measurements and a two wire (2W)
configuration for high resistances (up to 100 G$\Omega$ - using a
Keithley 2400 SourceMeter) or for a constant voltage measurement. We
also performed pulsed current-voltage measurements (I-V
characteristics), by generating a single square pulse of increasing
voltage (up to 10 V) for 20 ms to 2 s (using an Agilent 33250A 80MHz
Function/Arbitrary Waveform Generator) and determining the current
by measuring the voltage in a calibrated resistance using an
oscilloscope or directly with the SourceMeter for longer pulses.
Temperature was measured by a small diode thermometer well thermally
anchored to the sample. Most of the electrical transport experiments
were performed without applying an external magnetic field. When
small magnetic fields were applied ($H \sim$ 100 Oe) in order to
compare with magnetic measurements, we followed a zero
field-cooled-warming procedure (ZFCW). Magnetization measurements,
ZFCW and field-cooled-cooling (FCC) modes, were also performed as a
function of temperature using a Lake-Shore 7400 VSM magnetometer to
determine the magnetic response of these single crystals. Besides,
in order to estimate the variation of the ferromagnetic volume
fraction upon the application of an increasing electric field,
magnetization measurements at a fixed temperature ($T_0$) were also
performed simultaneously with a 2W $\rho(T_0,V)$ measurement.

\section{RESULTS AND DISCUSSION}

\begin{figure}
\centerline{\includegraphics[angle=0,scale=0.4]{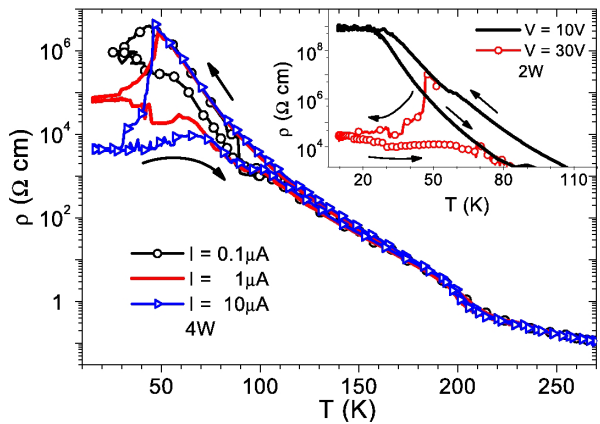}}
\vspace{-0mm} \caption{Resistivity (4W, current controlled) of LPCMO
as a function of temperature for various applied currents. The
temperature evolution is indicated by arrows. The inset shows the
2W, voltage controlled resistivity.} \vspace{0mm} \label{fig:RvsTYV}
\end{figure}

Fig.~\ref{fig:RvsTYV} and its inset show the resistivity of a LPCMO
(y $\sim$ 0.4) single crystal as a function of temperature for
different constant currents and voltages (in the inset). For a low
voltage range ($V \leq$ 10 V), the sample remains insulating down to
low temperatures. A saturation is observed most probably due to the
fact that the resistivity of the sample is beyond our measurement
capability. When higher voltages are applied (10 V $\leq V \leq$ 100
V), a metal-insulator like transition can be observed. A resistive
drop of more than four orders of magnitude is obtained for the
highest voltages as well as a characteristic temperature hysteresis
for this system~\cite{Uehara99}. A similar behavior is observed for
the current-controlled experiment.

\begin{figure}
\centerline{\includegraphics[angle=0,scale=0.3]{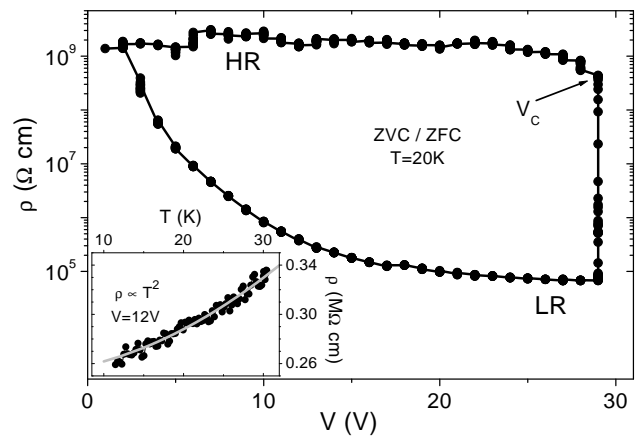}}
\vspace{-0mm} \caption{2W ZVC Resistivity of LPCMO as a function of
the applied voltage at $T$=20 K. The system evolves from a high
resistance (HR) to a low resistance (LR) regime, after applying a
voltage $V \geq V_c$. The inset shows a $T^2$ dependence of the
resistivity in the LR regime.} \vspace{0mm} \label{fig:RvsV20}
\end{figure}

Fig.~\ref{fig:RvsV20} shows the voltage dependence of the
resistivity ($\rho(V)$) at a fixed temperature ($\sim$20 K), for a
sample cooled in zero applied voltage (ZVC) and in zero magnetic
field (ZFC). A very high and constant resistivity (HR) is measured
as far as the voltage is increased up to a temperature dependent
critical value ($V_c(T)$), where a drop of more than four orders of
magnitude is observed. This critical or breakdown voltage depends
linearly with the increasing rate of the applied voltage. At this
point, a time evolution of the resistivity is observed, which can be
easily noticed by the reduction of the resistivity at a constant
voltage $V_c(T)$. When the voltage is then decreased, the sample
follows a different and more conducting path (LR), eventually
reaching the same initial resistivity for low voltages. If the
voltage is increased again, memory effects are observed as the
sample remains in this LR path, showing a voltage reversible
behavior and a $T^2$ temperature dependence, as can be observed in
the inset of Fig.~\ref{fig:RvsV20}. Similar results were obtained
for temperatures $T \leq$ 50 K.

\begin{figure}
\centerline{\includegraphics[angle=0,scale=0.3]{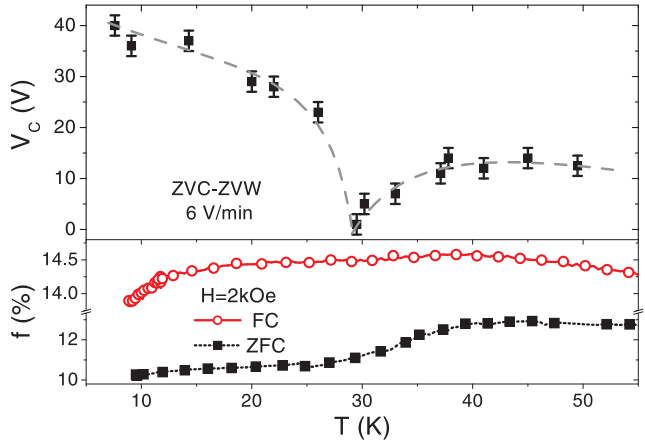}}
\vspace{-0mm} \caption{Temperature dependence of the critical
voltage ($V_c$) for a zero-voltage-cooled experiment (upper panel).
The lower panel shows the ferromagnetic volume fraction $f$ as a
function of temperature in the zero-field-cooled and the
field-cooled-cooling modes.} \vspace{0mm} \label{fig:VcyMT}
\end{figure}

The temperature dependence of $V_c(T)$ for a zero voltage
cooled-zero voltage warming experiment (ZVC-ZVW) can be observed in
the upper panel of Fig.~\ref{fig:VcyMT}, where the voltage was
increased for each case at a 6V/min rate. In the lower panel, the
temperature dependence of the ferromagnetic volume fraction ($f$) is
plotted for comparison. It should be noted that $V_c \rightarrow 0$
for a temperature $\sim$ 30 K, which is lower than the one
corresponding to the maximum of $f$ ($\sim$ 40 K) and can be related
to a change in the slope of the ZFC magnetization curve.

\begin{figure}
\centerline{\includegraphics[angle=0,scale=0.28]{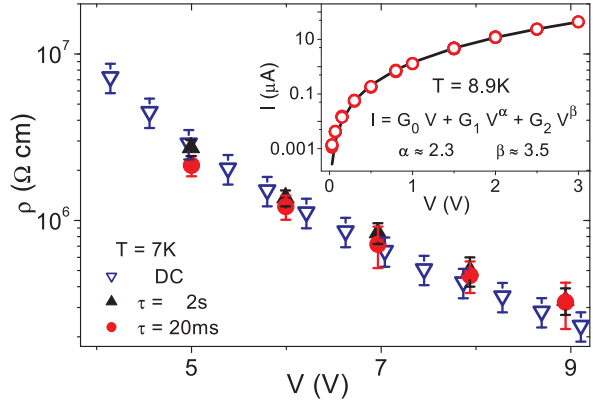}}
\vspace{-0mm} \caption{$\rho(V)$ curves measured in the LR regime
applying single square pulses of different time width ($\tau$=20 and
2000 ms) compared to DC measurements. The inset shows the particular
$I(V)$ characteristics of the LR regime.} \vspace{0mm}
\label{fig:pulsed}
\end{figure}

As when the sample reaches the LR regime the current considerably
increases, flowing principally in a metallic percolation path, an
additional Joule dissipation may occur. In this case, the shape of
the highest voltage parts of these $\rho(V)$ curves can be modified
by overheating. To rule out this possibility, in particular for $V
\leq V_c$, we performed pulsed $\rho(V)$ measurements in the LR
regime. As it is shown in Fig.~\ref{fig:pulsed}, no time dependence
can be observed for pulsed and DC voltages up to $V <$ 10 V, which
can be interpreted as a lack of an overheating contribution to the
shape of these $\rho(V)$ characteristics.

In these conditions the obtained $\rho$ shows a decrease with
increasing $V$. This non-linear behavior in the LR regime also shows
a particular power law dependence of the current ($I$) with voltage
($I \sim aV+bV^{7/3}+cV^{7/2}$, as shown in the inset of
Fig.~\ref{fig:pulsed}). This particular behavior was previously
observed analyzing the electrical transport properties of grain
boundaries in manganites~\cite{Klein99,Hofener00,Gross00,Chashka01}
and was interpreted within the framework described by the
Glazman-Matveev (GM) theory~\cite{Glazman88}, which characterizes
the case of multi-step tunneling of localized states within the
grain boundary between two ferromagnetic conducting regions. The
tunneling nature of the conducting process derived from our results
indicates that the conducting percolation path established in the LR
regime is not single connected.

\begin{figure}
\vspace{-10mm}
\centerline{\includegraphics[angle=0,scale=0.33]{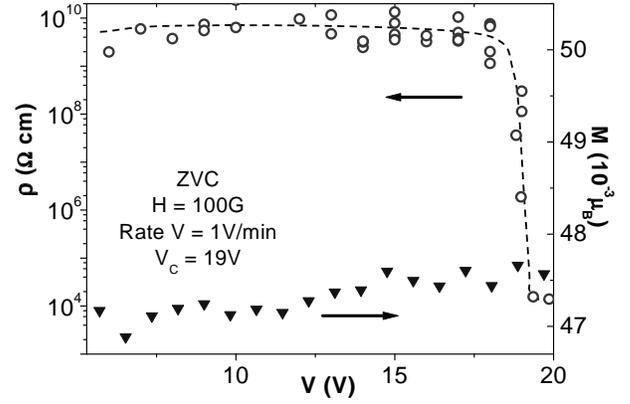}}
\vspace{-4mm} \caption{Simultaneous measurements of the voltage
dependence of the resistivity and the magnetization in a
zero-voltage-cooled and zero-field-cooled experiment, performed at
$T \sim$ 26 K. The dashed line is a guide to the eye.} \vspace{0mm}
\label{fig:RMV}
\end{figure}

In order to gain insight on the evolution of the CD-F volume
fraction ($f$) upon the application of a voltage $V>V_c$, we have
measured simultaneously $\rho(V)$ and $M(V)$ at low magnetic fields
($H$=100 Oe). Our results, obtained at $T \sim$ 26 K, are shown in
Fig.~\ref{fig:RMV}. Surprisingly, although our magnetic measurement
sensitivity let us to detect variations in $f$ lower than 1\%, no
appreciable changes on the magnetization ($M$) are detected when
voltage produces the large resistivity drop at $V_c$. This lack of
increase of the ferromagnetic volume fraction points toward the
development of a filamentary percolation path, generated upon the
increase of the applied voltage. This scenario is similar to the one
usually observed during the dielectric breakdown of an insulator,
where conducting defects increase with increasing voltage to finally
produce a percolative path.

Another way to reach the same conclusion, namely that a 1D CD-F
percolative path is established when the resistivity drop is
observed (instead of a 2D or 3D network with an $f$ higher than a
critical percolation value $f_c$), comes from the analysis of our
resistivity measurements by using the general effective medium (GEM)
equations (Eq.~\ref{eq:GEM}) developed by
McLachlan~\cite{McLachlan87} to describe the electric conductivity
of a binary mixture of conducting and insulator materials:

\begin{equation}
\label{eq:GEM} f
\frac{(\sigma_M^{1/t}-\sigma_E^{1/t})}{(\sigma_M^{1/t}+A\sigma_E^{1/t})}+
(1-f)
\frac{(\sigma_I^{1/t}-\sigma_E^{1/t})}{(\sigma_I^{1/t}+A\sigma_E^{1/t})}=0
,
\end{equation}
\noindent where $f$ is the volume fraction of the CD-F domains,
$\sigma_M$ and $\sigma_I$ the conductivities of the metallic and
insulating phases, respectively. $\sigma_E$ is the effective
conductivity that we measure, $t$ and $f_c$ are a critical exponent
and the percolation threshold, respectively. $A=\frac{1-f_c}{f_c}$,
and assuming that $\sigma_M$(T)=$\sigma$(T) of x=0 and
$\sigma_I$(T)=$\sigma$(T) of
x=0.625, the values of $f(V)$ can be calculated.\\

Within this GEM framework, both 2D or 3D percolation scenarios
($t_{3D}$=2 or $t_{2D}$=1 to 1.4 , f$_c$=0.17) yield to a variation
of more than a 4\% in $f$ when the sample passes from the HR to the
LR regime at $V_c(T)$ (T$<$33 K).~\cite{garba04b} This variation of
the ferromagnetic fraction should be easily noticed in the M(V)
measurements, which was not the case, confirming the filamentary
spatial distribution of the percolating path.

The PS scenario at low temperatures for this prototypical manganite,
which can be depicted as small and isolated CD regions in a CO
matrix, recalls, as we mentioned previously, the physics observed
for dielectric breakdown in metal loaded dielectrics.~\cite{Beale88}
These materials were built as a mixture of a small fraction of a
conducting component in an insulating matrix, which is an artificial
representation of the intrinsic PS regime observed in manganites.
Their dielectric breakdown field ($V_{bd}$), related to a series of
microscopic failures, depends on the conducting volume fraction, as
described by an empirical relation of the form:

\begin{equation}
\label{eq:Vbd} \frac{1}{V_{bd}(T)}= A(f) + (f_c-f)^{-\nu} ln(L),
\end{equation}
\noindent where $A$ is a parameter, $f$ the volume fraction of the
CD-F domains, $f_c$ the percolation threshold, $\nu$ a critical
exponent ($\sim$ 0.85 in 3D)\cite{Essam80} and $L$ a linear
dimension of the sample.

\begin{figure}
\vspace{5mm}
\centerline{\includegraphics[angle=0,scale=0.27]{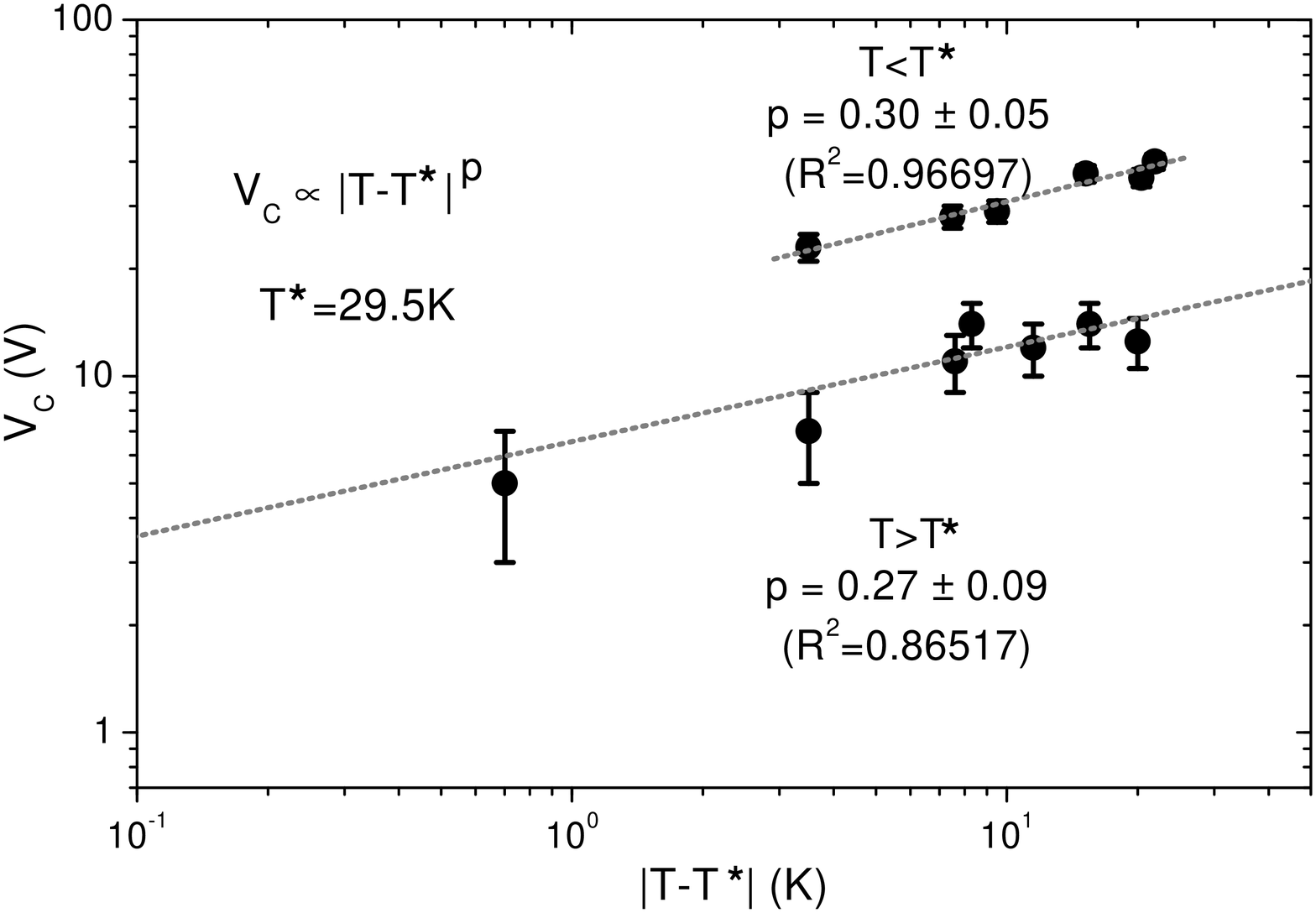}}
\vspace{-0mm} \caption{Critical electric field as a function of the
absolute relative temperature $t=|T-T^{\star}|$. A critical exponent
$p \simeq$ 0.3 and a $T^{\star} \simeq$ 29.5 K are obtained from the
best fits represented by the dotted lines.} \vspace{0mm}
\label{fig:VcvsTmTc}
\end{figure}

According to Eq.~\ref{eq:Vbd} and from the temperature dependence of
the ferromagnetic volume fraction $f$ obtained from the ZFC
magnetization measurement, it is easy to notice that the expected
$V_{bd}$ should essentially follow the soft temperature variation of
the ZFC magnetization and will not show, at any case, the quite
unusual temperature dependence of $V_c$ (showed in
Fig.~\ref{fig:VcyMT}). This fact reveals the existence of another
process, probably associated with a dielectric constant increase
that may enhance considerably the local electric field ($V_{bd} =
\epsilon(T) V_c(T)$). Indeed, $V_c(T)$ follows a critical behavior,
described by a $|T-T^{\star}|^p$ dependence, as can be observed in
Fig.~\ref{fig:VcvsTmTc}. The best fitting parameters correspond to
the temperature $T^{\star} \simeq$ 29.5 K and to the critical
exponent $p \simeq$ 0.3.

Neglecting the small temperature dependence of $V_{bd}$, as
$\epsilon \sim V_c^{-1}$, the obtained critical exponent may
correspond to a divergence of the dielectric constant with a
critical exponent $p_{\epsilon} = -p \simeq$ -0.3. This particular
divergence of the dielectric constant $\epsilon$ was previously
observed for binary liquid mixtures near the consolute critical
point and can be associated with the Maxwell-Wagner
effect~\cite{Thoen89,Hamelin90}. This effect, related to the low
frequency dielectric constant, results from the accumulation of
conducting charges at the interface in the boundary between two
phases of different conductivity. The effect is amplified near the
consolute point, as the system, that behaves as homogenous far below
this critical temperature, becomes heterogeneous (phase separated)
for temperatures above $T^{\star}$, with the occurrence of
large-size phase fluctuations in a critical region. Remarkably, this
framework seems to match very well with the PS scenario of
manganites, where the CD-F and the CO-AF phases coexist in a
determined temperature range. In our particular case, below
$T^{\star}$ part of the CD-F and CO-AF phases become miscible, in
accordance to the observed ferromagnetic volume fraction reduction,
and show large fluctuations between phase-separated and fully mixed
(paramagnetic) volumes in a critical region near $T^{\star}$. These
CD-F volume fluctuations favor the formation of a percolation path
when applying an external electric field, thus reducing the value
needed to reach the critical field ($V_c(T^{\star}) \rightarrow 0$
V).

Recently, Ghivelder et al.~\cite{Ghivelder05}, showed for this
particular manganite, the existence of a boundary between dynamic
and frozen PS effects, with a peak in the magnetic viscosity at a
temperature pretty much close to $T^{\star}$. By similarity with the
binary liquid mixtures, our results may indicate that the reported
change in the PS dynamics can be explained by the appearance of this
consolute-like point that should produce anomalies in many static
and dynamic properties at $T^{\star}$, as a consequence of the large
fluctuations developed near that critical
temperature~\cite{Hohenberg77}.

\section{CONCLUSIONS}
We have shown that the low-temperature electrical resistivity of
LPCMO single crystals is very sensitive to the magnitude of the
applied electric field. The resistivity of a
zero-electric-field-cooled sample can be varied more than four
orders of magnitude by increasing the exciting voltage over a
temperature dependent critical value ($V_c(T)$). Our simultaneous
electric and magnetic measurements indicate that a filamentary
percolation path is established when the resistivity evolves
irreversibly from a high to a low resistivity regime. The non-linear
$I-V$ characteristics obtained point out to the formation of a
non-single connected path where multi-tunneling processes determine
the electric transport properties of this regime. The presence of a
consolute-like point at $ T^{\star}$ was established by analyzing
the critical behavior of $V_c$, that we associate with a
Maxwell-Wagner effect. Thus, the presence of a consolute critical
point sheds light on the way that PS occurs, determining the
particular behavior of the properties observed for this compound
near the critical temperature.

\section{ACKNOWLEDGEMENTS}
Partial financial support from a UBACyT (X198), ANPCyT PICT 03-13517
and CONICET PIP 5609 grants is acknowledged. Work at Rutgers was
supported by the NSF-DMR-0520471 grant. GG acknowledges a
scholarship from CONICET of Argentina. We are indebted to G.
Pasquini for a critical reading of the manuscript and to D.
Gim\'enez, D. Melgarejo, C. Chiliotte and E. P\'erez-Wodtke for
their technical assistance.


\end{document}